# Computational Simulations of Solvation Force and Squeeze Out of Dodecane Chain Molecules in Atomic Force Microscopy

Rong-Guang Xu, Yuan Xiang, and Yongsheng Leng[*]

*Department of Mechanical and Aerospace Engineering*

*The George Washington University*

*Washington, DC 20052, USA*

There is a growing interest since the 1990s to understand the squeezing and shear behaviors of liquid films at nanometer scale by the atomic force microscope (AFM) measurement. We carry out all-atom contact-mode AFM simulations in a liquid-vapor molecular dynamics ensemble to investigate the solvation force oscillation and squeeze out mechanisms of a confined linear dodecane fluid between a gold AFM tip and a mica substrate. Solvation force oscillations are found to be associated with the layering transition of liquid film and unstable jumps of AFM tip position. Detailed structural analyses and molecular animations show that the local permeation of chain molecules and the squeeze out of molecules near the edge of contact promote the layering transition under compression. The confinement-induced slow down dynamics is manifested by the decrease in diffusivity and increase in rotational relaxation times. However, the persistent diffusive behavior of dodecane chain molecules even in the single-monolayer film is attributed to the chain sliding motions, given the fact that substantial vacancy space exists in the film due to thermal fluctuations.

[*] Corresponding author, Tel.: 202-994-5964; e-mail: leng@gwu.edu.

# I. INTRODUCTION

Much of our present understanding of the thermodynamic and mechanical properties of nonpolar fluids under nanometers confinement comes from extensive experimental studies during the past decades by the surface force apparatus (SFA) or surface force balance (SFB) instruments.[1-6] In these studies a small amount of fluid was put into a cross-cylinder confined geometry between two *macroscopic* mica surfaces. The lateral dimension of the cross-cylinder contact geometry could be as large as many microns.[6] It was generally acknowledged that the confined liquids behave remarkably different from those in the bulk. However, the nature of the squeeze out of nanoconfined fluids is still a controversial debate.[6-13] This fundamental question has clear applications in many fields such as nanomanufacturing and nanotribology.

In a much smaller lateral dimension of *nanometers* range, atomic force microscope (AFM) has been used to probe the squeezing and shear behaviors of nanoconfined liquid films.[14-27] This technique opens a new way of calibrating liquid films in a much smaller length scale, providing measurements of localized interaction forces in a nanoscale volume of the confined material. The nanoscale contact in all three dimensions in AFM enables a full-scale computational molecular dynamics (MD) simulation to probe the molecular interactions and squeezing phenomenon in AFM. One of the advantages of MD simulations of AFM in liquid media is that the tip geometry and roughness are well defined, avoiding some uncertainties and complications involved in AFM tip-sample contacts in many AFM experiments.

Computational molecular simulations of nanoconfined fluids in SFA are still facing many challenges due to the large micron size contacts.[6] For the squeeze out of liquid films in a contact area in SFA experiments, it has been suggested that confined liquid molecules need to travel "many microns" during the layering transition,[28, 29] therefore a hydrodynamic model is suitable to describe the lateral motions of confined molecules. We point out that this may not be the real situation in SFA since our recent computational simulation of a simple nonpolar fluid under nanoconfinement clearly showed that the squeeze out front dynamics was not controlled by the coherent sliding of liquid monolayer behind the front,[30] rather, the motion of fluid particles was much slower than the front speed. These particles simply underwent local permeations, resulting in layering transition. The same mechanism was also proposed for the outward squeezing observed in SFA[31], suggesting that permeation, rather than the large scale coherent sliding of monolayer, controls the mass transport.

In the present paper we focus on computational MD simulations of solvation force and squeeze out of a nonpolar fluid in AFM. We recently carried out the first MD simulation work on the solvation force oscillation of a model liquid,[32] the octamethylcyclotetrasiloxane (OMCTS) $[Si(CH_3)_2O]_4$ in an AFM contact geometry. This is a widely studied simple nonpolar (*globular*) liquid system in many SFA and AFM experiments. We found that solvation force oscillations of OMCTS extend to $n = 8$ layers and the solidlike nanoconfined film holds at $n = 2$ layers.[32] Here, we turn our attention to a *long-chain* nonpolar molecular fluid, an alkane fluid. Recent AFM solvation force measurements showed that many nonpolar liquids exhibit oscillatory forces at the nanoscale, irrespective of molecular geometry.[29] However, the information of structural evolution of nanoconfined fluids during squeeze out is still not available, needing molecular simulations to explore the detailed molecular mechanism of layering transitions.

It should be noted that unlike previous molecular simulation work of nanoconfined fluids in SFA experiments,[33-38] our current simulation study applies a driven dynamic model in a liquid-vapor molecular ensemble to mimic the squeeze out of the nanoconfined fluid in an AFM tip-substrate contact.[32] We intend to use this simulation method to explore the squeezing out mechanism of dodecane ($C_{12}H_{26}$) under nanoscale contact in AFM. We choose dodecane as the model alkane fluid because it was experimentally studied previously in AFM solvation force[16, 29] and SFA viscosity[39] measurements for the confined fluid. Previous computational efforts were also done for the same fluid through MD simulations,[33-38] but largely focused on molecular packing structure, shear behavior and nano-rheology of the confined fluid between two molecularly smooth solid surfaces.

In the simulation work of OMCTS,[32] we developed a new molecular model that includes internal degrees of freedom for this *globular* molecule. A major difference between OMCTS and dodecane *chain* molecule is that OMCTS is quite rigid, while dodecane is quite flexible with free rotations around all C-C bonds. This difference in molecular geometry and rigidity may lead to different molecular packing structures and thus the solvation force and squeeze out behaviors.

The rest of the paper is organized as follows. In section II, the molecular model and simulation method will be described in detail. In section III, we present detailed MD simulation results and discussion, followed by our summary in section IV.

## II. MOLECULAR MODELS AND SIMULATION METHOD

## A. The molecular model of dodecane

Figure 1 shows the molecular structure of dodecane, a flexible chain molecule with the chain length about 1.8 nm and the diameter of each $CH_2$ segment of 0.5 nm. It has been demonstrated that both the homogeneous[40, 41] and heterogeneous[42-45] united-atom (UA) intermolecular potential models are inadequate for accurately simulating transport properties such as shear viscosity, self-diffusion coefficient, and internal relaxation times of long chain and branched alkanes, particularly at higher densities such as the conditions under confinement. It was generally believed that viscosity in UA model is always underestimated and diffusion coefficient is overestimated.[46] The molecular geometry and dynamics of dodecane can be well described by the all-atom optimized potentials for liquid simulations (OPLS-AA) force field,[47] which is suitable to describe the energy and dynamics of hydrocarbon molecules. The potential energy associated with dodecane includes intramolecular bond stretching, bond angle bending, dihedral angle torsion, and nonbonded Lennard-Jones (LJ) terms, as well as the Coulomb electrostatic interaction term. The nonbonded LJ and electrostatic interactions with 1-4 scaling factor can be expressed as[47]

$$E_{ij} = \left\{ \frac{q_i q_j}{r_{ij}^2} + 4\varepsilon_{ij} \left[ \left( \frac{\sigma_{ij}}{r_{ij}} \right)^{12} - \left( \frac{\sigma_{ij}}{r_{ij}} \right)^{6} \right] \right\} f_{ij} \quad (1)$$

where $\varepsilon_{ij}$ and $\sigma_{ij}$ are the well depth and characteristic distance between two atomic species. For hydrocarbon molecules of OPLS-AA model, we have $\sigma_C$ = 3.5 Å, $\varepsilon_c$ = 0.066 kcal/mol, and $\sigma_H$ = 2.5 Å, $\varepsilon_H$ = 0.030 kcal/mol. Furthermore, standard combining rules are used, where $\sigma_{ij} = (\sigma_i \sigma_j)^{1/2}$ and $\varepsilon_{ij} = (\varepsilon_i \varepsilon_j)^{1/2}$. The factor $f_{ij}$ in eq. (1) is 1-4 scaling factor (SF), which is SF = 0.5 if $i$ and $j$ are 1-4 interaction pair within one molecule and SF = 1.0 otherwise.

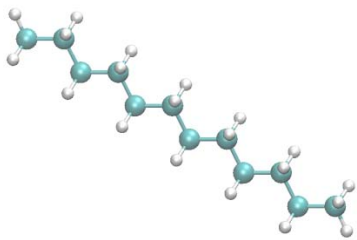

**Figure 1.** Conformation of a dodecane chain molecule. Oxygen and hydrogen atoms are shown in cyan and white, respectively.

However, it was shown that the original OPLS-AA parameters for hydrocarbons developed for short alkanes cannot properly reproduce the bulk properties of long-chain alkanes with carbon number equal or exceeding 12.[48] To illustrate this, we show in Fig. 2(a) the artificial crystallization of a bulk liquid of dodecane in a *NPT* MD ensemble under ambient condition ($T$ = 298 K and $P$ = 1 bar). The problem lies in the scaling factor SF = 0.5 for the 1-4 interaction, which has a substantial effect on the properties of n-dodecane molecules. This observation is also consistent with early studies by Ye et al.,[48] from which a suggestion was made that a smaller scaling factor as low as SF = 0.0 for the 1-4 interaction should be adopted for long-chain alkanes. Using SF = 0.0 for the 1-4 interaction (Fig. 2 (b)), together with the original OPLS-AA parameters with a modified torsion potential,[49] we can reproduce the key results such as the bulk density and self-diffusion coefficient of dodecane, as compared with the work by Ye et al.[48] (see Table I).

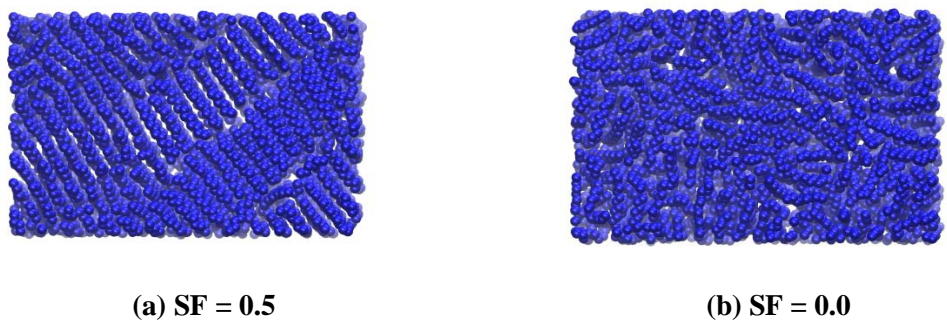

**(a) SF = 0.5**    **(b) SF = 0.0**

**Figure 2.** The effect of scaling factor (SF) on the packing structure of dodecane at $T$ = 298 K and $P$ = 1 bar in the bulk. A total of 3.5 ns MD equilibrium run is performed for 512 dodecane molecules in a *NPT* ensemble: (a) SF = 0.5; and (b) SF = 0.0.

TABLE I. Comparisons of the bulk density and diffusion coefficient of dodecane obtained from experiment and MD simulations.

|  | bulk density (g/cm$^3$) | diffusion coefficient ($10^{-10}$ m$^2$/s) |
| --- | --- | --- |
| X. Ye et al. [48] | 0.755 | 5.75 |
| Experiment [a] | 0.745 | 8.71 |
| This work | 0.738 | 7.70 |

[a] See the cited references therein.[48]

## B. The AFM simulation in liquid medium

Following our previous AFM simulation work in OMCTS,[32] we choose muscovite mica (chemical formula $K_2Al_4(Al,Si_3)_2O_{20}(OH)_4$) as a molecularly smooth substrate. We use the CLAYFF force field[50] to describe the dynamics of mica with its very bottom potassium ions $K^+$ being fixed. AFM tip is modeled by a gold tip to reflect the surface chemistry of the tip according to AFM experimental procedures.[22, 23, 51] We use the embedded atom method (EAM) potential to describe the dynamics of gold atoms in the AFM tip. The EAM model is an updated version developed by the force matching method.[52] When a nonpolar fluid such as dodecane is confined between an AFM gold tip and a mica substrate, the tip-dodecane and mica-dodecane molecular interactions are largely dominated by dispersive forces. Therefore, we use the LJ type interaction as shown in eq. (1) to describe theses molecular interactions. LJ parameters for mica and gold are available in the CLAYFF[50] and UFF[53] force fields. From these, other parameters for interactions between different species are obtained according to the simple combining rule discussed in II A.

The detailed sketch of the AFM tip-substrate contact in liquid medium is shown in Fig. 3(a). Here we use a driving spring model to represent the elasticity of the AFM cantilever in three dimensions. Only the top two rigid-layer atoms of the gold tip are connected to the driving springs. A liquid-vapor molecular ensemble is used to maintain a vapor pressure of the ambient condition.[54, 55] The AFM gold tip and mica substrate are completely immersed in the liquid phase to avoid capillary effect, with periodic boundary conditions applied along the three directions. The AFM tip is modeled as a spherical tip truncated by a flat bottom exposing (111) facet. The tip radius is about 4.5 nm. The radius of the flat bottom surface is about 2.7 nm. The dimensions of the simulation box along the *x*-, *y*-, and *z*-directions are 70, 14.4, and 10.5 nm, respectively. The molecular system includes 4800 dodecane molecules (totally 182,400 atoms), 6624 gold atoms, and a two-layer mica sheet containing 24 × 16 × 2 mica unit cells (42 atoms per unit cell with a total of 32,256 atoms in mica substrate).

Figure 3 (b) shows the molecular equilibrium configuration of the gold tip-mica substrate in dodecane after more than 4 ns MD equilibrium run. Very stable liquid-vapor boundaries of dodecane on the two sides of simulation system are clearly seen.

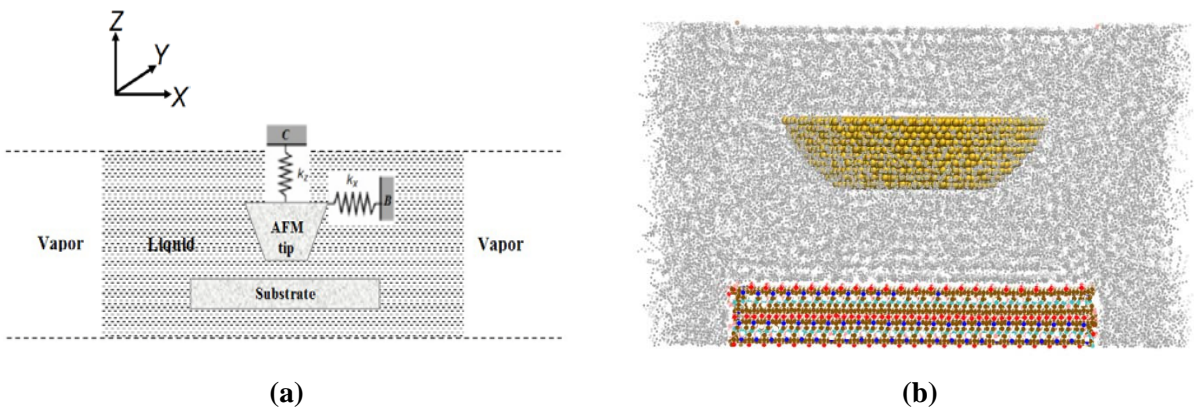

**Figure 3.** (a) Computational model of the AFM simulation in a liquid-vapor molecular ensemble; (b) A snapshot of the gold tip-mica substrate in dodecane fluid at thermodynamic equilibrium.

In AFM solvation force simulations, we select the normal spring constant $k_z = 8$ N/m. This is a typical value applied in AFM liquid force measurements.[16, 23, 51, 56, 57] The gold tip is also constrained in the $x$- and $y$- directions by lateral springs ($k_{x,y} = 215$ N/m, see Fig. 3(a)). The driving support C is pushed downward to compress the spring, making the AFM tip slowly approaching the mica substrate in dodecane. The compression speed is $v = 1$ m/s. The temperature of the molecular system is controlled at 298 K by a Nose-Hoover thermostat and the time step in MD simulation is 1 fs. The cutoff distance of 11.0 Å is used for the LJ interactions, while the long-range electrostatic interactions between charged particles are treated by the particle-particle-particle-mesh (PPPM) method.[58]

## III. RESULTS AND DISCUSSION

### A. Solvation force and layering transition

Figure 4 shows the simulated force-distance curves as the AFM gold tip approaches the mica substrate in dodecane. The force curve is qualitatively similar to those observed for OMCTS and argon liquids,[30, 32] i.e., the force oscillates as a function of gap separation, alternating between maxima and minima with increasingly pronounced repulsive force peaks as the gap separation is decreased. Three force peaks are clearly seen, starting from a distance at about 20 Å. The oscillatory periodicity of ~ 5 Å corresponds to the diameter of a fully stretched linear dodecane, or the size of a $CH_2$ segment. This indicates that the layered structures of dodecane are formed between the gold tip and mica substrate. The progressively increasing force peaks during normal compression shows that the last three- or two-layer film becomes more difficult to be squeezed out. Associated with the solvation force oscillations, molecular animations show that the minima in the force curve correspond to the newly-formed layered structures, while the maxima in the force curve correspond to the critical stage prior to the layering transition. Molecular animations further show that during the normal approach of the gold tip towards the mica substrate, the $n \to n-1$ layering transition is accomplished by a sudden downward jump of the gold tip. In our simulation, the proportions of time for $n = 3 \to 2$ and $2 \to 1$ unstable jumps (with steep positive slopes) are about 35% and 6% of the total time for squeezing $n = 3$ and $n = 2$ films, respectively. The remaining time is largely occupied by stable pushing of the film, as shown by the blue arrows in Fig. 4 (a) with negative slopes. These unstable jumps of the gold tip are also directly shown in Fig. 4(b).

The instability during layering transition is well-known upon the spring stiffness is smaller than the gradient of the solvation force, i.e., $\partial F/\partial D > k_z$. Usually these unstable regions cannot be accessible by soft AFM cantilevers, but can be probed by *dynamic* AFM with very stiff springs. It is only at the last one or two layers of dodecane that strong repulsive force peaks are observed. Squeezing out these last two or even one layer of dodecane would need to overcome a large energy barrier. At larger distances the relatively small force peaks indicate that the layered structure of dodecane is not compact. Also noted is that during the normal compression of dodecane to the last two layers, there is no obvious elastic deformation associated with the mica substrate. This is quite different from the SFA experiment in which the flattening of curved mica surfaces is typically observed due to the large contact area at micrometer scales.

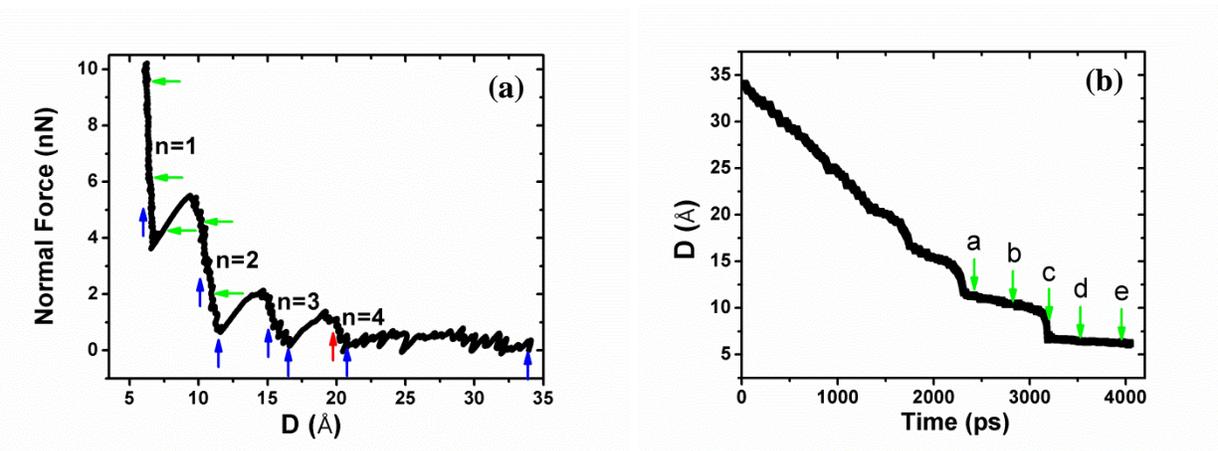

**Figure 4.** (a) The solvation force versus gap distance and (b) the AFM tip position versus time variations obtained from AFM simulations in dodecane. Blue arrows in (a) represent stable stages at which long time MD relaxations will not result in force decays. The red arrow shows the unstable position. Green arrows in both panels correspond to the snapshots a), b), c), d), and e) in Fig. 5 during the squeeze out process.

Figure 5 shows the detailed configuration changes of dodecane in the contact region during the $n = 2 \to 1$ layering transition. Molecular animation I (see supplementary material) shows the detailed layering transition. Note that only about 0.05 ns out of the total of 0.85 ns compression of the $n = 2$ film (about 6% of compression time, see Fig. 4(b) the tip displacement jump from b to c) was taken to complete this unstable dynamic transition, which was associated with the structural change from panel (b) to panel (c) in Fig. 5. Here, dodecane chain molecules shown in red and blue colors in Fig. 5 represent the initial top and bottom layers, respectively. As the unstable transition proceeds, the confined molecules simply undergo local permeation during the layering transition, pushing the chain molecules near the edge of contact leaving the confined region. This detailed squeeze out process is dramatically different from the claim that a *collective motion* of one monolayer proceeds during the layering transition.[28, 29] In a recent SFA experimental study on the squeeze out of OMCTS,[31] it was suggested that the squeeze out front propagation is not controlled by large scale flow in the confined film. Our simulation results support the idea that permeation, rather than the large scale coherent sliding of monolayer, controls the mass transport. In Fig. 5, the last snapshot clearly shows that the final equilibrium structure of $n = 1$ layer is made up of mixed red and blue dodecane chain molecules, supporting the local permeation explanation.

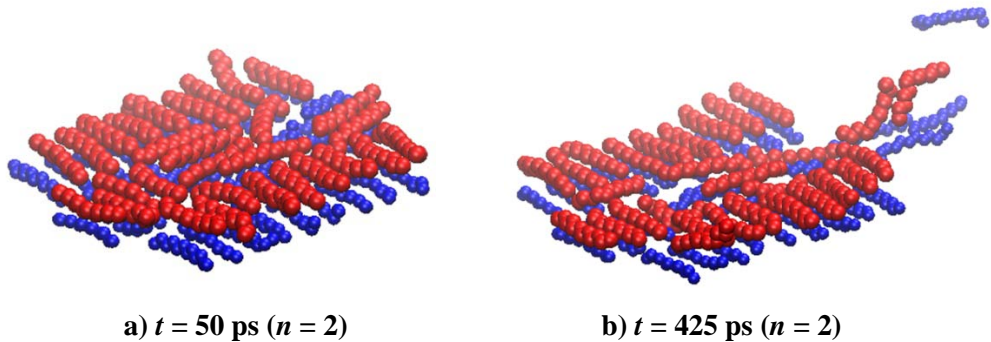

a) $t = 50$ ps ($n = 2$)   b) $t = 425$ ps ($n = 2$)

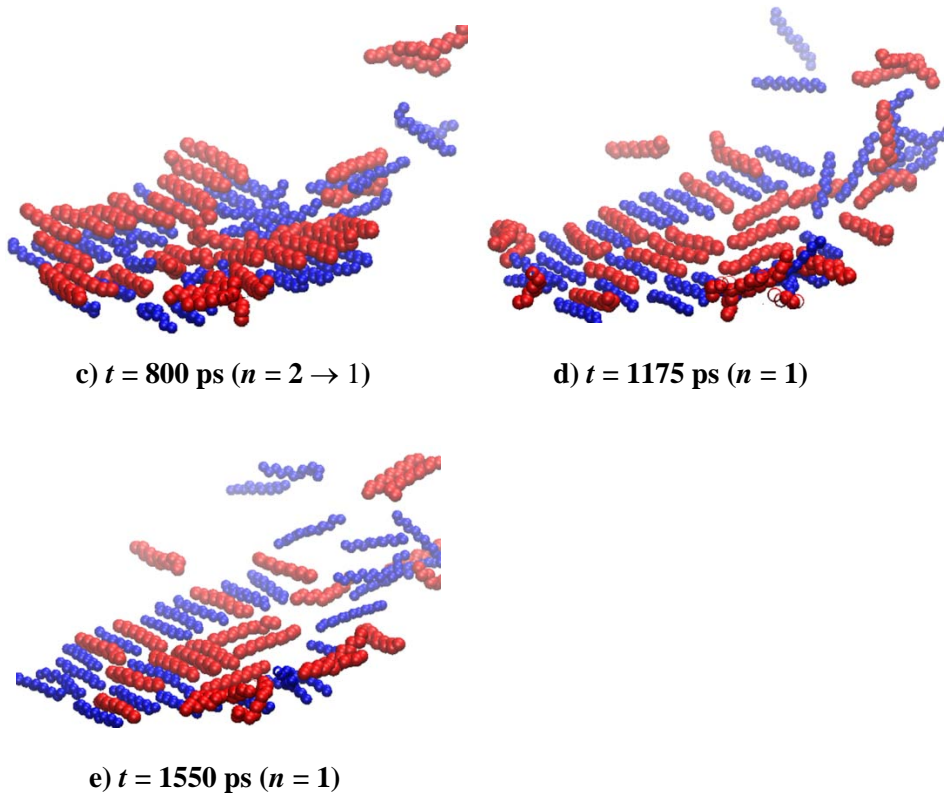

**c)** $t = 800$ ps ($n = 2 \rightarrow 1$)     **d)** $t = 1175$ ps ($n = 1$)

**e)** $t = 1550$ ps ($n = 1$)

**Figure 5.** The dynamic progression of the squeeze out process of dodecane chain molecules confined between AFM tip and substrate. Red: initial top layer. Blue: initial bottom layer

**B. Dodecane density distributions**

In order to investigate the layered structures of dodecane at different distances, we further perform MD relaxations at each specific distance for 2 ns MD run by holding the driving spring stationary. These distances are marked by blue arrows in Figure 4, at which no force decays are observed during relaxations. The red arrow at $n = 4$ layer distance turns out unstable and therefore is no longer considered for further structural analysis. Figure 6 shows the density distributions of dodecane across the gap between the gold tip and mica surface. It should be noted that in the central region the small and wide density peaks indicate that the layered structure of dodecane is not compact. The overall asymmetric density distribution is due to the asymmetric tip-dodecane and mica-dodecane contacts. Moreover, for $n = 3$ and $n = 2$ films, two density distribution curves are shown for each case, revealing the dependence of density distribution on the extent of film compression.

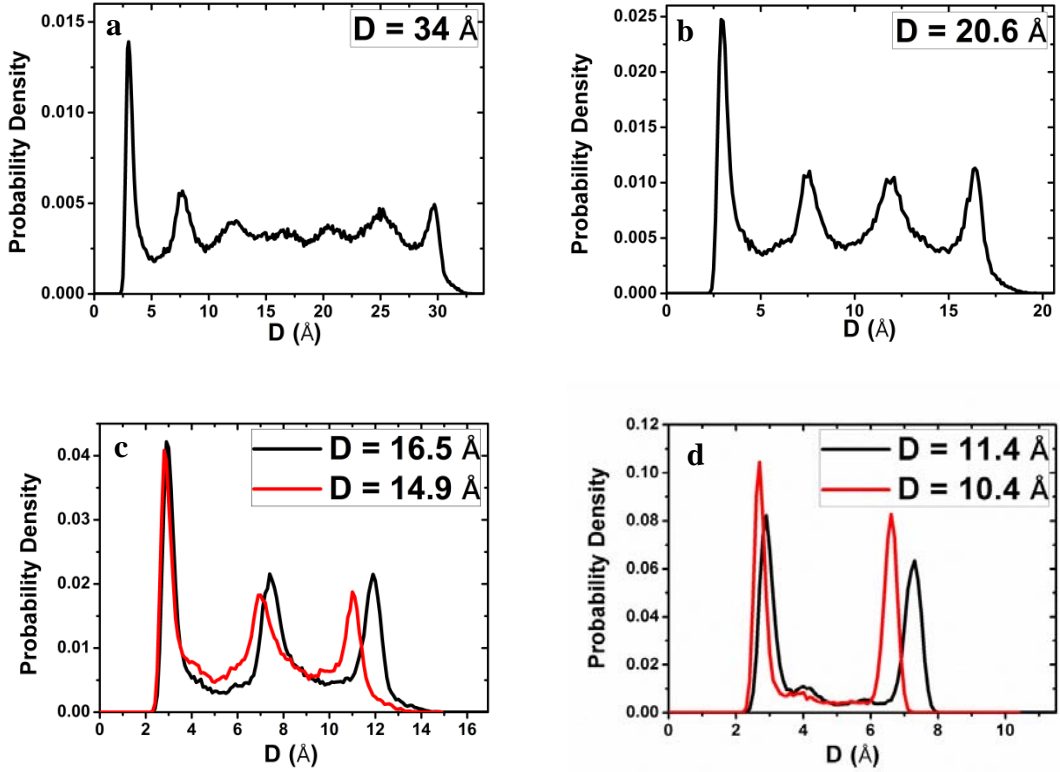

**Figure 6.** Dodecane density distributions (a - d) across the gap between a gold tip and mica surface in different layered films.

## C. Dodecane local orientation and orientation pair distribution function

To quantitatively monitor the overall structural changes associated with the confined dodecane molecules during the squeeze out process, we further calculate the local orientation orders (LOO) of chain molecules and the molecular orientation pair distribution function (MOPDF) at different layered films. The local orientation order (LOO) of the confined film is defined as

$$LOO = \frac{1}{2}\langle 3\langle \cos^2 \theta \rangle - 1 \rangle, \qquad (2)$$

where $\theta$ denotes the angle between the end-to-end vector of dodecane backbones and the $z$-normal direction. The triangle brackets denote the average over time and dodecane molecules in the confined region. LOO will take a value of 1.0, 0.0, or −0.5 for the dodecane molecules being parallel, random or perpendicular to the $z$-direction. The LOO values corresponding to different layers are shown in Fig. 7. It is quite obvious that the orientation order parameter monotonically decreases from 0.0 to −0.5 as the gap distance is decreased. In the bulk phase, the dodecane molecules have an isotropic, random backbone orientation (LOO = 0.0), while in the $n = 1$ film, LOO = −0.5, signifying a layered structure in which all the dodecane molecules in the confined region are perpendicular to the $z$-axis, or parallel to the mica substrate.

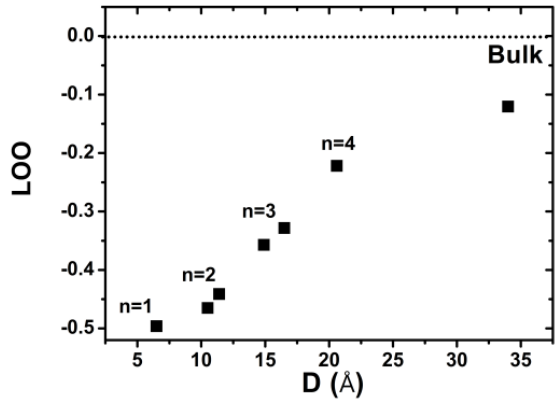

**Figure 7.** Layer-by-layer local orientation order (LOO) of confined dodecane molecules.

The MOPDF of the confined dodecane film is defined as

$$f(\theta) = \frac{1}{n}\left\langle \sum_i \sum_{j \neq i} \delta(\cos\theta_{ij} - \cos(\theta)) \right\rangle, \quad (3)$$

where $\theta_{ij}$ is the angle between the end-to-end vectors of molecules $i$ and $j$. The triangle bracket denotes an ensemble average and the summation runs through all dodecane molecules, $n$, in the confined region. The MOPDFs corresponding to different gap distances are shown in Figure 8. At larger distances, the confined dodecane is in the liquidlike state, resulting in relatively flat MOPDFs. As the gap distance is gradually decreased, the dodecane molecules are self-organized into a layered structure with their orientations largely parallel with each other ($\cos\theta = \pm 1$), which can be seen by increasing values at the two ends of MOPDFs in Fig. 8. A typical molecular configuration of $n = 3$ film can be seen in Fig. 9 at a gap distance of D = 14.9 Å. However, the dodecane packing structures of $n = 2$ and $n = 1$ films (corresponding to D = 10.4 Å and D = 6.4 Å, respectively) also adopt the perpendicular structure to which a small peak around $\cos\theta = 0$ emerges (see Fig. 8). This perpendicular alignment molecular packing structure is also clearly seen in Fig. 5.

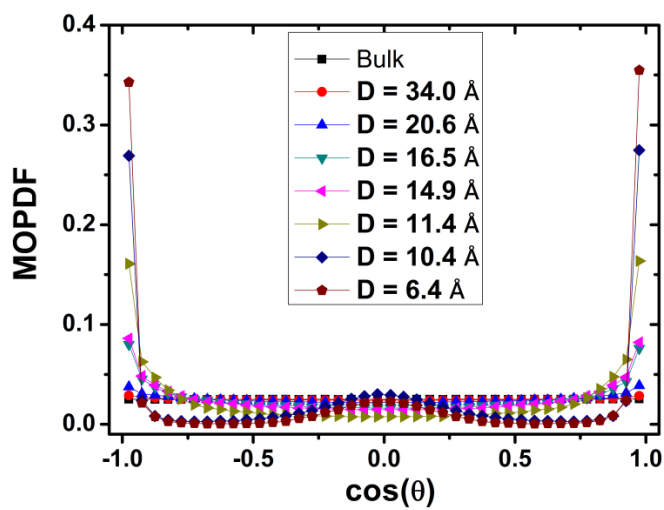

**Figure 8.** The molecular orientation pair distribution function (MOPDF). The molecular parallel orientation corresponds to $\cos\theta = \pm1$, while the perpendicular orientation corresponds to $\cos\theta = 0.0$.

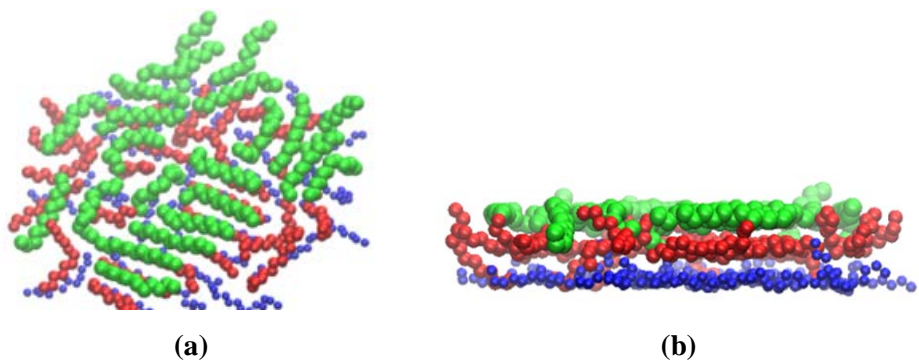

(a)          (b)

**Figure 9.** Molecular configuration of dodecane confined between the gold tip and mica surface in $n = 3$ layered film (D = 14.9 Å). (a) top view and (b) side view.

### D. Dynamic properties of dodecane molecules in confined films

We now consider the detailed dynamic behaviors of confined dodecane molecules between the gold tip and mica subtrate. Molecular diffusion coefficient and rotation correlation time are two major quantities in this investigation.

*Diffusion Behavior* The translational motion of dodecane molecules can be well described by the self-diffusion coefficient, *D*, determined by the Einstein relation

$$\langle |r(t) - r(0)|^2 \rangle = 2dDt \qquad (4)$$

where $r(t)$ is the position of the center of mass of dodecane molecules at time $t$, $d$ is the dimensionality of the space in which the diffusion is considered. Here, $d = 2$ because we only consider the diffusion in the lateral direction. The mean square displacement (MSD) $<|r(t)-r(0)|^2>$ is calculated over all the dodecane molecules in the confined region and the time origin average is also considered.

Figure 10 shows the MSD curves of dodecane molecules in the confined region. From the data we can calculate self-diffusion coefficients of dodecane in different layers. The MSD curve for the bulk at 298 K and 1 bar pressure is also shown in the figure. The calculated bulk diffusion coefficient of dodecane from *NPT* MD simulation is about $7.70\times10^{-10}$ m$^2$/s, slightly less than the experimental value of $8.71\times10^{-10}$ m$^2$/s (see Table I). Table II shows the calculated diffusion coefficients of dodecane in different films. Even for the $n = 3$ layered film (Fig. 9), the in-plane diffusion coefficient of dodecane is still about 40% of the bulk value. When the gap distance is further decreased to two- and one-layer films, the diffusion coefficient of dodecane continues to decrease, but within one order of magnitude at the maximum (for $n = 1$ layer). This somewhat surprising result, especially for $n = 1$ layered film, seems not supportive to the claim of "solidified" film under extreme confinement. To further understand how the molecular diffusion of dodecane could happen under this extraordinary condition, in Fig. 11 we show two time frames of molecular configuration of dodecane in $n = 1$ layer. We find that molecules highlighted by ellipses undergo "molecular sliding" due to inevitable molecular vacancies in the film. These vacancies are generated by the in-and-out molecular exchanges of dodecane near the edge of contact region, as well as by the thermal fluctuations. Molecular animation II (see supplementary material) shows the detailed molecular sliding and molecular exchange events at the contact interface, which ultimately contribute to the increase in MSD.

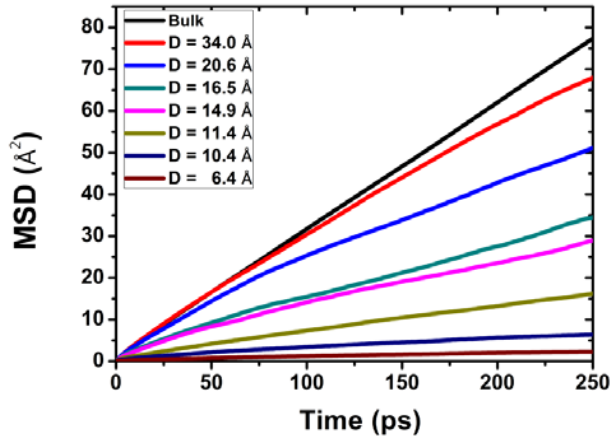

**Figure 10.** The mean square displacements (MSDs) of dodecane chain molecules in the confined region as a function of time. The MSD curve for the bulk fluid is also shown in the figure for comparison.

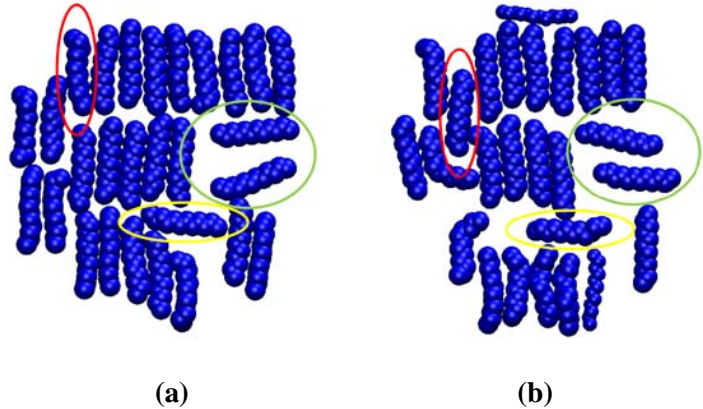

**Figure 11.** Molecular configurations of dodecane in $n = 1$ film at two time frames: a) $t = 0$ ns and b) $t = 0.5$ ns. Dodecane molecules circled by ellipses contribute to nonzero MSD.

TABLE II. Diffusion coefficients ($D$) of dodecane in the bulk and in different confined films between the gold tip and mica substrate.

| gap (Å) | ∞ (bulk) | 34.0 | 20.6 | 16.5 | 14.9 | 11.4 | 10.4 | 6.40 |
|---|---|---|---|---|---|---|---|---|
| $D$ ($10^{-10}$ m$^2$/s) | 7.70 | 6.81 | 4.89 | 3.21 | 2.88 | 1.56 | 0.60 | 0.21 |

***Rotation Behavior*** It is well known that confined fluids have a spectrum of relaxation times. Here we investigate the time variations of the rotational autocorrelation function of dodecane, mainly focusing on the rotational dynamics of the end-to-end vector of dodecane, which is represented by a unit vector **S**. The rotational autocorrelation function of **S** is defined as the first rank Legendre polynomial in NMR experiment[59], given by

$$P_{1,S}(t) = \langle S(t) \cdot S(0) \rangle = \langle \cos(\theta_S(t)) \rangle \tag{5}$$

where $\theta(t)$ is the angle between the vector **S** at time $t_0$ and that at time $t_0 + t$. The time average runs over all the dodecane molecules in the confined region. The correlation functions $P_1(t)$ can be well fitted by the Kohlrausch-Williams-Watts (KWW) stretched exponential function[60]

$$P_1(t) = \exp\left[-(t/\tau)^\gamma\right] \tag{6}$$

in which $\tau$ and $\gamma$ are two fitting parameters. The time integral of $P_1(t)$ is analytic, giving a rotation correlation time as

$$\tau_c = \int_0^\infty P_1(t)dt = \frac{\tau}{\gamma}\Gamma(\frac{1}{\gamma}) \tag{7}$$

where $\Gamma$ is the gamma function. The MD simulation results are shown in Fig. 12. Table III summarizes the fitted parameters and correlation times of dodecane in the bulk and in different confined films. As the confined film is thinning, the rotational correlation time of dodecane is increased. However, this increase in time is not dramatic until $n = 1$ layer is reached.

TABLE III. Rotation correlation times of dodecane in different confined films.

|  | $\gamma$ | $\tau$(ps) | $\tau_C$(ps) |
|---|---|---|---|
| bulk | 0.8247 | 103.145 | 114.433 |
| D = 34.0 Å | 0.9183 | 173.018 | 180.102 |
| D = 20.6 Å (n = 4) | 0.7472 | 173.632 | 207.378 |
| D = 16.5 Å (n = 3) | 0.6285 | 412.496 | 585.759 |
| D = 14.9 Å (n = 3) | 0.6407 | 423.021 | 587.477 |
| D = 11.4 Å (n = 2) | 0.5895 | 1926.025 | 2966.411 |
| D = 10.4 Å (n = 2) | 0.5617 | 3009.883 | 4948.785 |
| D = 6.4 Å (n = 1) | 0.2411 | $2.592 \times 10^8$ | $7.789 \times 10^9$ |

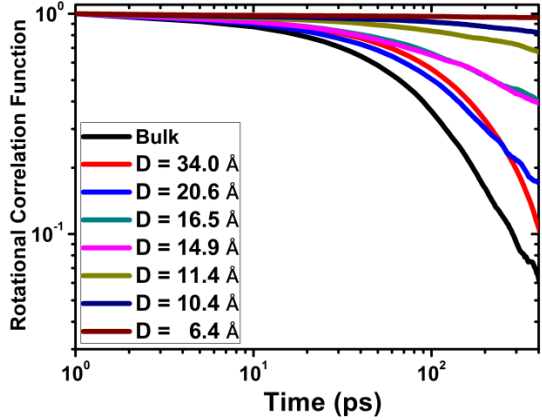

**Figure 12.** The variations of rotational correlation functions of dodecane molecules in the confined region as a function of time.

## IV. SUMMARY AND FURTHER DISCUSSION

In the present work, realistic all-atom contact-mode AFM simulations have been performed to explore the solvation force and squeeze out mechanisms of a confined linear dodecane fluid between a gold AFM tip and a mica substrate. We employ the OPLS-AA force field for dodecane that can correctly describe its bulk properties in liquid phase with the scaling factor being set to SF = 0 for the 1-4 intra-molecular interaction. We especially use a driving spring model in a liquid-vapor molecular dynamics ensemble to mimic the AFM force measurement under ambient condition. For the nanoscale contact in AFM which is much smaller than the contact area in the SFA or SFB, the confined dodecane chain molecules can still self-organize into a layered structure. The solvation force-distance curve obtained from MD simulation exhibits force oscillations with a period of dodecane molecular diameters. The unstable force/tip position jumps associated with the layering transition are well captured by the driving spring model, which are similar to AFM experimental observations. More importantly, the dynamic evolutions of molecular packing structure during this layering transition clearly show that instead of a collective motion of one monolayer, the local permeation of confined molecules and the squeeze out of the molecules near the edge of contact contribute to the overall squeeze out process. The decrease in diffusivity and increase in rotational relaxation times clearly indicate the confinement-induced slow down dynamics. However, even in the single-monolayer film the notable diffusivity of dodecane molecules in the form of chain sliding is attributed to the inevitable vacancies in the layered structure, which are induced by constant in-and-out molecular exchanges of dodecane near the edge of contact area, and also by the thermal fluctuations in the confined layer.

It is therefore interesting to understand how the effect of lateral dimension of contact or confinement influences the layering transition and diffusivity of molecular fluids under compression, in particular for the micron contact in the SFA or SFB experiments. In our previous molecular simulation work of a simple nonpolar fluid (argon) under confinement,[30, 61] we showed the first evidence of inward/outward squeeze out front and vacancy diffusion in the solidified film. For more realistic simple nonpolar fluids such as OMCTS and cyclohexane ($C_6H_{12}$), and dodecane chain molecules, molecular simulations of dynamic squeeze out and sliding friction of these complex fluids under large-area confinement will shed more light on the force oscillation and stick-slip friction observed in SFA or SFB experiments. Moreover, dynamic oscillations of mica surfaces, similar to our very recent work of the first dynamic AFM simulation in liquid,[62] may reveal the new relationship between the contact stiffness and damping of the confined film in large contact area.

**Supplementary Material**

See supplementary material for the complete molecular animation I (n = 2 → 1 layering transition) and II (molecular sliding diffusion).

**ACKNOWLEDGMENTS**

This work is supported by the National Science Foundation (NSF 1149704) and the National Energy Research Scientific Computing Center (NERSC).